\documentclass[preprint]{aastex62}
\usepackage{natbib}
\submitjournal{ApJ}

\begin{document}

\title{X-ray Irradiation of the Giant Planet Orbiting the T Tauri Star TAP 26}

\correspondingauthor{Stephen L. Skinner}
\email{stephen.skinner@colorado.edu}

\author{Stephen L. Skinner}
\affiliation{Center for Astrophysics and
Space Astronomy (CASA), Univ. of Colorado,
Boulder, CO, USA 80309-0389}

\author{Manuel  G\"{u}del}
\affiliation{Dept. of Astrophysics, Univ. of Vienna,
T\"{u}rkenschanzstr. 17,  A-1180 Vienna, Austria}

\newcommand{\ltsimeq}{\raisebox{-0.6ex}{$\,\stackrel{\raisebox{-.2ex}%
{$\textstyle<$}}{\sim}\,$}}
\newcommand{\gtsimeq}{\raisebox{-0.6ex}{$\,\stackrel{\raisebox{-.2ex}%
{$\textstyle>$}}{\sim}\,$}}
\begin{abstract}
\small{
We present new Chandra X-ray observations of TAP 26, a $\approx$17 Myr old 
magnetically-active weak-lined T Tauri star that has been reported to host a 
massive planet in a $\approx$10.8 day orbit. At a separation of $a$ = 0.097 AU 
the planet will be exposed to intense X-ray and UV radiation from the star. 
The first observation caught the star in a state of elevated X-ray emission
with variability on a timescale of a few hours and an X-ray temperature
kT$_{x}$ $\approx$ 2 - 4 keV. Two subsequent observations 5-10 days later
showed slow variability and a lower X-ray flux and temperature (kT$_{x}$ $\approx$ 1 keV).
We characterize the X-ray emission and estimate the X-ray ionization and 
heating rates that will need to be incorporated into realistic models of
the planet's atmosphere.
}
\end{abstract}

\section{Introduction}
Observational studies of planet-hosting T Tauri stars (TTS) provide
crucial information on the environments in which planets form around
solar-like stars. At ages of a few Myr the high-energy X-ray and UV (XUV)
emission from TTS can be a thousand times more luminous than their
main-sequence descendants. This intense XUV radiation heats and 
ionizes gas in the inner disk and in the atmospheres of close-in
exoplanets, influencing atmospheric temperature, chemistry, and mass loss.
Thus the effcts of XUV radiation must be considered in models of
exoplanet formation, evolution, and even the survival of close-in planets.

Most exoplanet discoveries so far have been around old mature host
stars (ages $\gtsimeq$ 1 Gyr) but ongoing searches are now revealing
exoplanets around a few young TTS. The discovery of exoplanets orbiting
TTS a few million years old are important because they provide a snapshot
of conditions in the planet-forming environment during the early stages
of planet formation. One of the most compelling examples so far is
the weak-lined TTS PDS 70 for which two formative giant planets have
been directly imaged at deprojected separations of $\approx$20 - 40 AU
from the star (Keppler et al. 2018, 2019; Haffert et al. 2019).  
But at this separation the planets will be little affected by 
stellar XUV radiation (Skinner \& Audard 2022). In contrast, TTS
such as TAP 26 discussed here represent the oposite extreme
where the exoplanet orbits very close to the host star, thus exposing it
directly to the intense XUV radiation field. The X-ray emitting 
weak-lined TTS V830 Tau may also harbor a close-in 
massive planet and its properties are summarized below (Sec. 4.5).

\noindent {\em A Massive Planet  Orbiting TAP 26.}~
Periodic radial velocity variations in the weak-lined TTS TAP 26 (= V1069 Tau)
have been attributed to a giant planet V1069 Tau b (also called TAP 26 b) orbiting at a
close separation  $a$ = 0.097 au (Yu et al. 2017; hereafter Yu17),  about four
times smaller than Mercury's semi-major axis. The near-circular orbit has a period of
9 - 11 d with a most likely value P$_{orb}$ = 10.79 d.
The planet has a minimum mass M$_{p}$sin$i$ = 1.66 M$_{J}$ where
M$_{J}$ is Jupiter's mass. If the 
planet's orbital and equatorial planes are aligned at $i$ $\approx$ 55$^{\circ}$
then M$_{p}$ = 2.03$\pm$0.46 M$_{J}$ (Yu17).

The host star TAP 26 is a K7e type weak-lined TTS in the Taurus star-forming 
region at a {\em Gaia} DR3 parallax distance $d$ = 121.6$\pm$0.2 pc which we
adopt herein.  It is viewed
through low extinction A$_{\rm V}$ = 0.25-0.43 mag (Herczeg \& Hillenbrand 2014).
Stellar properties are given in Yu17 who assumed a distance of 147 pc and
derived a mass M$_{*}$ = 1.04$\pm$0.1 M$_{\odot}$, stellar luminosity
log (L$_{*}$/L$_{\odot}$) = $-$0.25$\pm$0.10, and an age of $\approx$17 Myr. The star 
is a rapid rotator with P$_{rot}$ = 0.7135 d (Grankin 2013) and is magnetically active (Yu17).

TAP 26 is a prominent X-ray source and has been detected by
ROSAT and the Einstein Observatory.
Being a $\sim$17 Myr old weak-lined TTS its disk has largely
dissipated and the exoplanet is directly exposed to intense
high-energy XUV stellar radiation. The 
XUV radiation is expected to heat the planet's atmosphere to a
maximum temperature of $\sim$10$^{4}$ K, well above
typical  equilibrium temperatures of a few hundred K to 
$\sim$2500 K (Fortney, Dawson, \& Komacek 2021).
As such, the TAP 26 system provides an
excellent laboratory to study a ``hot Jupiter''
in a harsh XUV environment that will
contribute to atmospheric heating and photoevaporation.

TAP 26 is one of the youngest T Tauri stars known so far to 
host a hot Jupiter. Since the mass of TAP 26 is nearly identical 
to the Sun it is a benchmark laboratory for constraining
giant planet formation around solar-like stars. The question of
how Jupiter-mass planets end up in orbits so close to their host 
stars is much debated and both {\em in situ} and migration models
have been proposed (see reviews by Dawson \& Johnson 2018;
Fortney et al. 2021). TAP 26 b migration scenarios were considered by Yu17.

\section{X-ray Observations} 

TAP 26 was detected in X-rays in the {\em ROSAT} All-Sky Survey (Neuh\"{a}user et al. 1995)
and in a {\em ROSAT} PSPC pointed observation (source 1WGA J0418.8$+$1723). 
It was also detected in two Einstein Imaging Proportional Counter (IPC)
observations (Feigelson et al. 1987) with a reported flux 
F$_{x}$(0.2-4 keV) = (3.9 - 4.6) $\times$ 10$^{-13}$ ergs cm$^{-2}$ s$^{-1}$.  

We observed TAP 26 with Chandra in December 2021 using the 
Advanced CCD Imaging Spectrometer (ACIS-S), as summarized in
Table 1. The 60 ks observation was obtained in three segments
over a timespan of $\approx$11 days as a result of Chandra's
operational constraints. 
Data were reduced using Chandra Interactive Analysis 
of Observations (CIAO vers. 4.14) software in combination with
CALDB  vers. 4.9.8 calibration data\footnote{For more information on
CIAO and CALDB see https://cxc.cfa.harvard.edu.}. 
Separate spectra and  X-ray light curves were generated for each 
observation using events within a circular region of radius 2$''$
centered on the source peak.
X-ray spectra and associated response matrix 
files (rmf) and  auxiliary response files (arf)  were 
extracted  using CIAO {\em specextract}. Energy-filtered light curves
were produced using CIAO {\em dmextract} in broad (0.3-8 keV),
soft (0.3-2 keV), and hard (2-8 keV) bands. 
CIAO {\em glvary} was used to check for source
variability and compute the probability of variability P$_{var}$.
Background and pileup were negligible. Spectra were analyzed using 
XSPEC vers. 12.10.1.

\begin{deluxetable}{llll}
\tabletypesize{\small}
\tablewidth{0pt}
\tablecaption{Summary of TAP 26 Chandra Observations}
\tablehead{
\colhead{Parameter} &
\colhead{}          &
\colhead{Observation}          &
\colhead{}          \\
}
\startdata
ObsId                       &  25693              & 25124                 & 26225                    \\
Start Date (2021)/Time (TT) &  Dec. 1/07:38       & Dec. 6/23:52          &  Dec. 11/13:40           \\
Stop  Date (2021)/Time (TT) &  Dec. 1/13:39       & Dec. 7/08:49          &  Dec. 11/17:02           \\
Start MJD (d)               &  59549.318          & 59554.995             &  59559.569               \\
Elapsed Time (ks)           &   21.665            & 32.211                &  12.078                  \\
Livetime (ks)\tablenotemark{b} & 17.989           & 27.227                &  9.456                  \\
$\phi_{rot}$\tablenotemark{c}  & 0.0 - 0.35       & 0.96 - 0.31           &  0.37 - 0.57            \\ 
\enddata
\tablenotetext{a}{Data were obtained using ACIS-S in faint timed event mode,
a frame time of 0.6 s, and the source positioned on CCD S3. 
The X-ray centroid position of TAP 26 from is (J2000)
R.A. = 04$^{\rm h}$18$^{\rm m}$51$^{\rm s}$.72, decl. = $+$17$^{\circ}$23$'$15$''$.6.
The {\em Gaia} DR3 position is 
R.A. = 04$^{\rm h}$18$^{\rm m}$51$^{\rm s}$.71, decl. = $+$17$^{\circ}$23$'$16$''$.4 .}
\tablenotetext{b}{Livetime is the time during which data were being collected 
and excludes operational and instrumental overheads such as CCD readout times.}
\tablenotetext{c}{Rotational phase is computed assuming P$_{rot}$ = 0.7135 d (Yu et al. 2017)
and $\phi_{rot}$ = 0 is arbitrarily defined to be the start of ObsId 25693.}
\end{deluxetable}



\clearpage

\section{Results}

\subsection{X-ray Light Curves and Spectra}

TAP 26 was detected as a variable X-ray source as summarized in
Table 2. The X-ray light curves are shown in Figure 1.
Hereafter, we refer to ObsId 25693 as high-state and
ObsIds 25124 and 26225 as low-state since the count rate
and flux were twice as high in ObsId 25693.

An abrupt rise and fall in broad-band count rate 
is clearly visible in the broad-band ObsId 25693 light curve
and the variability probability is P$_{var}$ = 0.99.
A slow decline in broad-band count rate also occurred in ObsId 25124
with the mean rate in the second half of the observation being 
$\approx$25\% less than the first half and P$_{var}$ = 087.
No significant variability was detected in the last and shorter 
observation ObsId 26225 (P$_{var}$ = 0.12).
The soft-band (0.3-2 keV) emission dominates the light curves in all observations.
But in ObsId 25693 the hardness ratio and hard-band (2-8 keV) count rate 
of 8.5$\pm$0.7 c ks$^{-1}$  were higher than the other two observations.
This reflects the higher plasma temperature that was measured 
in ObsId 25693 (Sec. 4.2).
ObsId 25124 was obtained about five days
after ObsId 25693 and at nearly the same rotational phase, but the
broad-band count rate and flux were a factor of two lower.
Thus the elevated emission detected in ObsId 25693 
is evidently not tied to the star's rotational phase (Table 1).

Spectra were fitted in XSPEC with absorbed ($tbabs$) one-temperature (1T) 
or two-temperature (2T)  $apec$ and $vapec$ (variable
abundances) thin plasma models. 
The absorption is low and not well-constrained by the data so
N$_{\rm H}$ was held fixed during fitting at 
N$_{\rm H}$ = 6.65 $\times$ 10$^{20}$ cm$^{-2}$ 
corresponding to A$_{\rm V}$ $\approx$ 0.35 mag
and the adopted conversion 
N$_{\rm H}$ (cm$^{-2}$) = 1.9$\pm$0.3 $\times$10$^{21}$$\cdot$A$_{\rm V}$
(Gorenstein 1975; Vuong et al. 2003).
The spectra obtained in ObsIds 25124 and 26225 when the 
count rate was lower (low-state) are nearly identical
and were thus fitted simultaneously in XSPEC to reduce
the uncertainties in best-fit parameters. A simple
isothermal 1T $apec$ model allowing the metallicity $Z$
to vary provides a very good fit of the low-state spectra and 
converges to kT $\approx$ 1 keV and a low  
metallicity $Z$ = 0.15$\pm$0.04 $Z_{\odot}$ (Table 2).
Since X-rays from weak-lined TTS are thought to arise in hot
magnetically-confined plasma analogous to the solar corona,
this metallicity reflects coronal abundances, not photospheric.

As shown in Figure 2, the high-state spectrum (ObsId 25693)
shows strong emission in the 0.9 - 1.2 keV range that is also
visible but weaker in the low-state spectrum (ObsId 25124).
Emission in this energy range is 
attributable to the Ne IX triplet (E$_{lab}$ = 0.92 keV, 
maximum line power at log T$_{max}$ = 6.6 K), Ne X (E$_{lab}$ = 1.02 
and 1.21 keV, log T$_{max}$ = 6.8 K), and a possible contribution from 
multiple closely-space Fe lines near 1 keV. 
The Si XIII triplet (1.86 keV, log T$_{max}$ = 7.0 K) is also 
clearly present in the high-state spectrum.
The high-state spectrum required a two-temperature model (2T $vapec$)  
to obtain a satisfactory fit and it is improved by
allowing the abundances of Ne and Fe to vary rather than 
just varying the metallicity $Z$. While varying the Ne abundance
it was constrained to be equal for both  temperature components and 
similarly for Fe.
The 2T $vapec$ fits converge to a low iron abundance 
Fe =  0.15$\pm$0.12 times solar but a Ne abundance 
above solar is required to reproduce the strong Ne
line emission (Table 2).  
The plasma was hotter in high-state with a characteristic
temperature weighted by the emission measure (EM) contributions of
the cool and hot components kT$_{wgtd}$ = 1.6 keV compared to
kT = 1.0 keV in low-state. The X-ray
luminosity was about 2.5 times higher in high-state.

\begin{deluxetable}{lcccclcc}
\tablewidth{0pt}
\tablecaption{Summary of TAP 26 X-ray Properties (Chandra ACIS-S)}
\tablehead{
           \colhead{ObsId (state)}      &
           \colhead{Rate}               &
           \colhead{Counts}             &
           \colhead{Hardness}           &
           \colhead{E$_{50}$}           &
           \colhead{kT}                 &
           \colhead{F$_{x,abs}$}        &
           \colhead{log L$_{x}$}        \\
           \colhead{}                   &
           \colhead{(c ks$^{-1}$)}      &
           \colhead{(c)}                &
           \colhead{Ratio}              &
           \colhead{(keV)}              &
           \colhead{(keV)}              &
           \colhead{(ergs cm$^{-2}$ s$^{-1}$)}  &
           \colhead{(ergs s$^{-1}$)}     }
\startdata
25693 (high)       & 44.4(v) & 798 & 0.26   & 1.49 &   1.62\tablenotemark{b} & 6.84$\pm$0.35e-13\tablenotemark{b}  & 30.20\tablenotemark{b} \\
25124 (low)        & 22.0(v) & 598 & 0.17   & 1.33 &   1.02\tablenotemark{c} & 2.77$\pm$0.10e-13\tablenotemark{c}  & 29.80\tablenotemark{c} \\
26225 (low)        & 21.5    & 203 & 0.16   & 1.28 &   1.02\tablenotemark{c} & 2.77$\pm$0.81e-13\tablenotemark{c}  & 29.80\tablenotemark{c} \\
\multicolumn{8}{c}{Time Partitioned Data\tablenotemark{d}}             \\
25693 (pre-flare)  & 37.9    & 248 & 0.28   & 1.56 &   2.82                  & 4.84$\pm$0.69e-13                   & 30.03 \\
25693 (flare)      & 56.1    & 420 & 0.26   & 1.49 &   2.03                  & 6.69$\pm$0.53e-13                   & 30.17 \\
25693 (post-flare) & 32.9    & 130 & 0.26   & 1.31 &   0.96                  & 3.21$\pm$0.26e-13                   & 29.86 \\
\enddata
\tablenotetext{a}{
Notes: 
Mean count rate (Rate = counts/livetime) where (v) denotes the rate was variable, counts, 
hardness ratio = counts(2-8 keV)/counts(0.3-8 keV), 
median event energy (E$_{50}$), plasma temperature (kT) in energy units weighted by
emission measure contribution for 2T models,
absorbed X-ray flux (F$_{x,abs}$$\pm$1$\sigma$),
and unabsorbed X-ray luminosity (L$_{x}$, at d = 121.6 pc).
Tabulated values were computed using events in the  0.3-8 keV range.}
\tablenotetext{b}{Based on a spectral fit of the ObsId 25693 spectrum rebinned to a minimum of
10 counts per bin using an absorbed 2T $vapec$ model of the
form N$_{\rm H}$*(kT$_{1}$ $+$ kT$_{2}$) with absorption (XSPEC $tbabs$) held fixed at  
N$_{\rm H}$ = 6.65e20 cm$^{-2}$ and allowing the abundances of Ne and Fe to vary
(Ne = 1.80$\pm$0.59, Fe = 0.15$\pm$0.12 $\times$ solar; referenced to the solar
abundances of Anders \& Grevesse 1989).
kT is the emission-measure (EM) weighted value of the two temperature components 
(kT$_{1}$ = 0.38$\pm$0.04, kT$_{2}$ = 2.91$\pm$0.47 keV, EM$_{1}$/EM$_{2}$ = 1.05,
 $\chi^2$/dof = 59.2/59 = 1.00, 1$\sigma$ errors).}
\tablenotetext{c}{Based on a simultaneous fit of the spectra for ObsIds 25124 and 26225
rebinned to a minimum of 10 counts per bin  using an absorbed 1T $apec$ plasma model of the form
N$_{\rm H}$*(kT) with absorption (XSPEC $tbabs$) held fixed at N$_{\rm H}$ = 6.65e20 cm$^{-2}$,
  kT = 1.02$\pm$0.04 keV, best-fit metallicity $Z$ = 0.15$\pm$0.04 $Z_{\odot}$, 
$\chi^2$/dof = 60.9/66 = 0.92, 1$\sigma$ errors).} 
\tablenotetext{d}{Based on fits of the ObsId 25693 spectra for the pre-flare (livetime 6551 s), 
flare (7488 s), and post-flare (3950 s) segments (Sec. 4.1). The value of P$_{var}$ was
not computed for individual time segments.}
\end{deluxetable}

\clearpage

\begin{figure}
\figurenum{1}
\includegraphics*[width=5.0cm,height=7.0cm,angle=-90]{f1t.eps} \\ 
\includegraphics*[width=5.0cm,height=7.0cm,angle=-90]{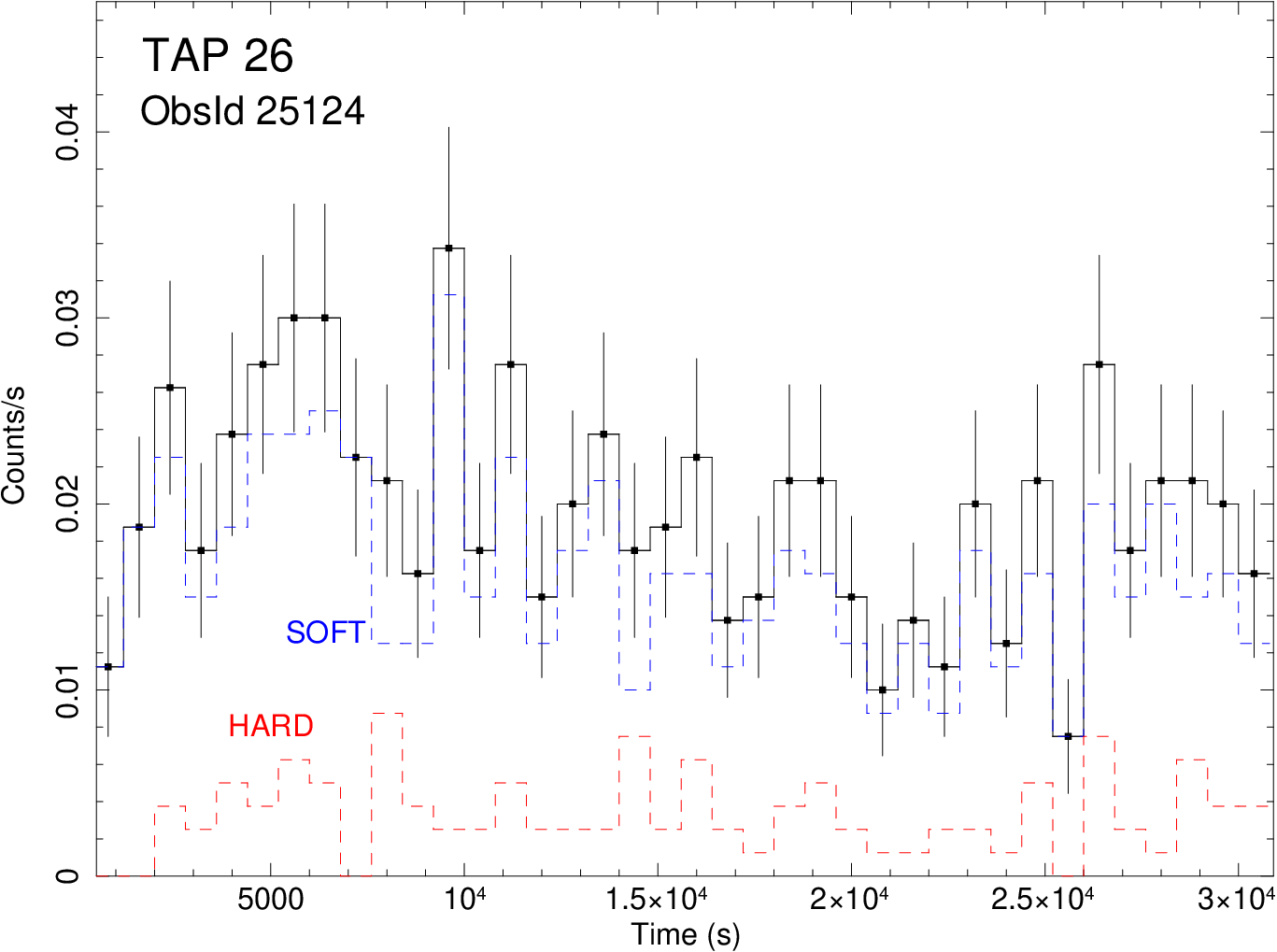} \\
\includegraphics*[width=5.0cm,height=7.0cm,angle=-90]{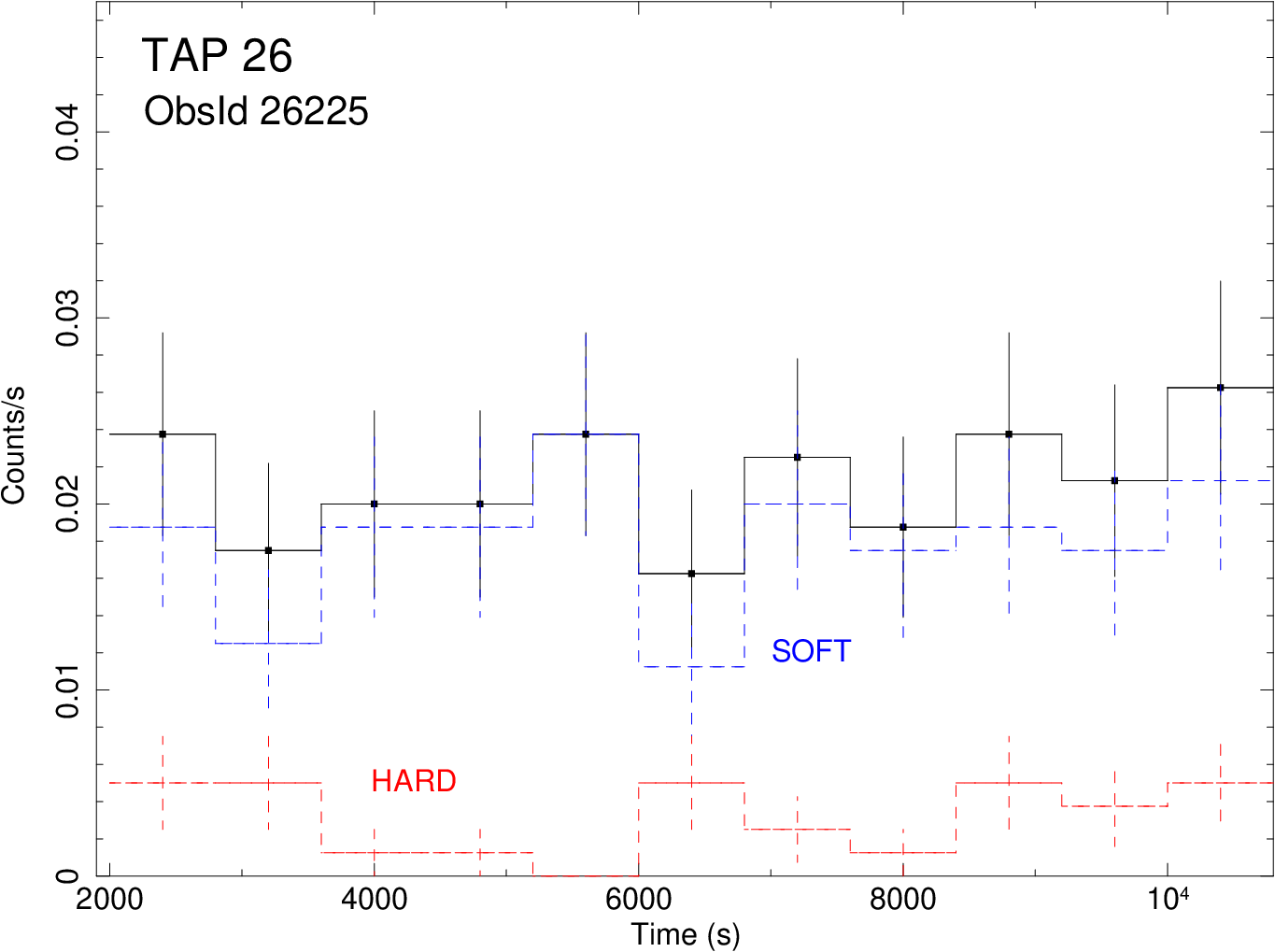} \\
\caption{{\em Chandra} ACIS-S light curves of TAP 26 binned at  800 s 
intervals in broad 0.3-8 keV (black), soft 0.3-2 keV (blue), and 
hard 2-8 keV (red) energy bands. 
Times are relative to start of each {\em Chandra} observation (Table 2).
For comparison, the solid horizontal solid line in the high-state
variable light curve of ObsId 25693 (top) shows the mean count 
rate (0.3-8 keV) during the short low-state observation ObsId 26225 (bottom).
}
\end{figure}

\begin{figure}
\figurenum{2}
\includegraphics*[height=3.5in,angle=-90]{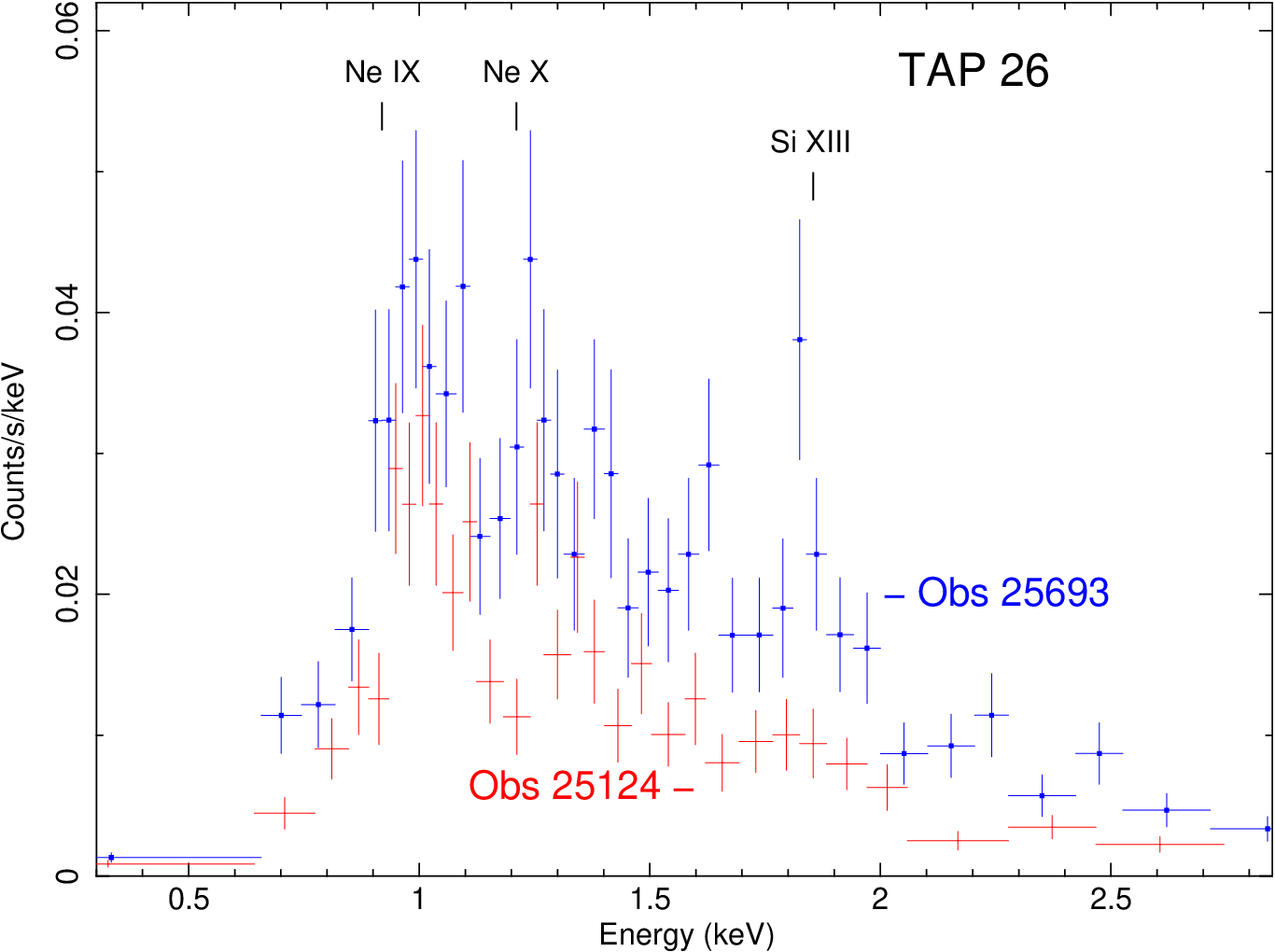} \\
\hspace{0.2in}
\includegraphics*[height=3.5in,angle=-90]{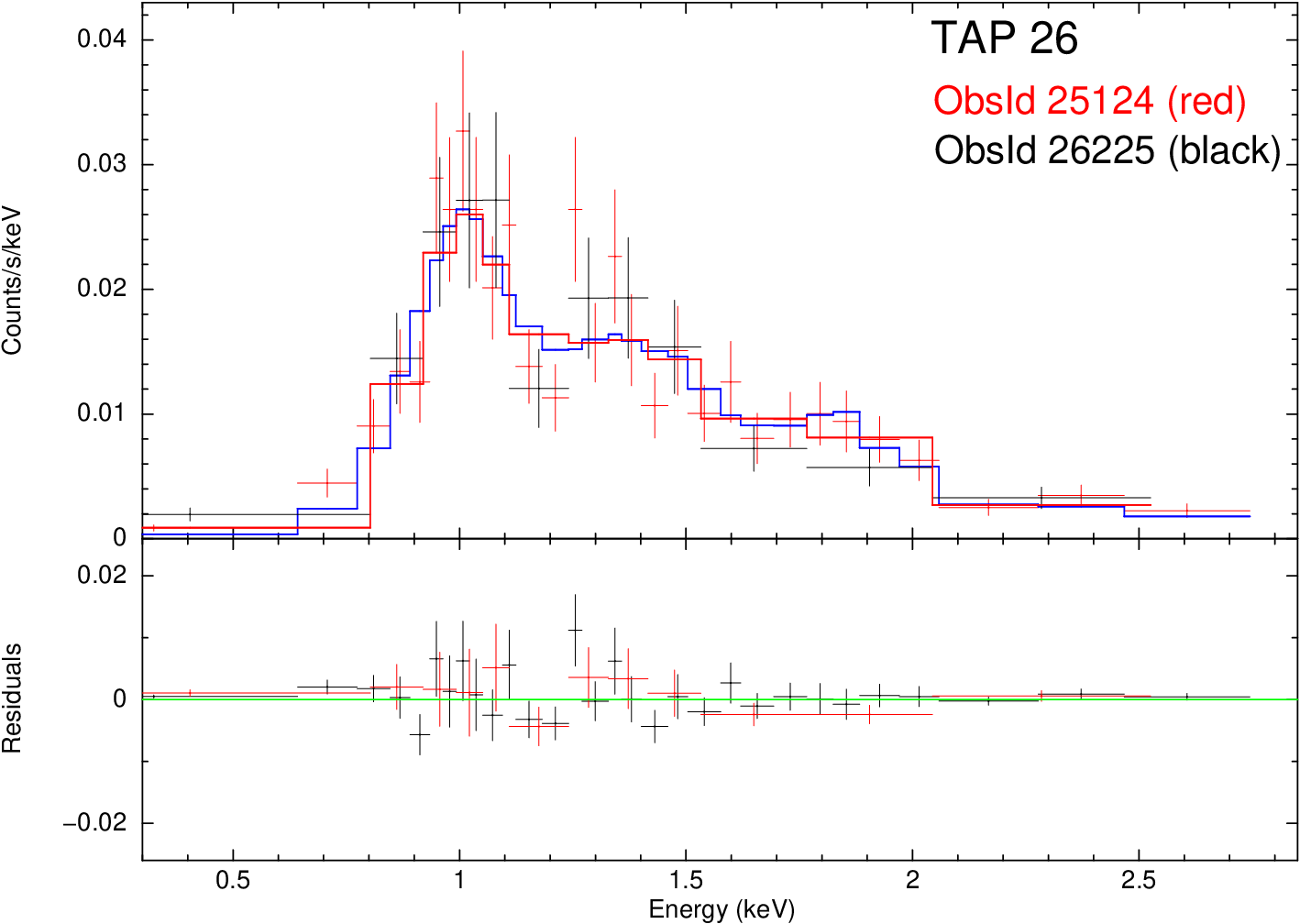}
\caption{Chandra ACIS-S spectra of TAP 26, rebinned for display.
Error bars are 1$\sigma$.
Top:~Overlay of high-state (ObsId 25693, blue)
and low-state (ObsId 25124, red).
A few candidate emission lines (or blends) are identified.
Bottom:~Overlay of TAP 26 low-state spectra for ObsId 25124 (red, 598 cts)
and 26225 (black, 203 cts).
The blue solid line is a simultaneous fit of the two spectra using
the 1T $apec$ model (kT = 1.02 keV) whose best-fit parameters are given 
in Table 2 notes. Fit residuals are shown at bottom. 
}
\end{figure}

\clearpage

\section{Discussion}

\subsection{X-ray Variability}

A comparison of the low and high-state measurements (Table 2) 
shows that the flux and X-ray luminosity of TAP 26 varied by a 
factor of $\approx$2.5 within a five day interval. The high-state
light curve for ObsId 25693 reveals short-term variability on a 
timescale of a few hours. An exponential fit of the declining X-ray 
light curve during the last half of ObsId 25693 after the count rate 
peaked gives an e-folding timescale of $\approx$9 ks.
In addition ObsId 25124 reveals a slow decline in count rate over 
$\approx$8 hours. For a comparison on a timescale of decades with 
the Einstein IPC we measured the Chandra fluxes in the 
IPC energy range and obtained
F$_{x}$(0.2-4 keV) = 2.72$\pm$0.14 $\times$ 10$^{-13}$ (low-state)
and 6.23$\pm$0.36 $\times$ 10$^{-13}$ (high-state)
ergs cm$^{-2}$ s$^{-1}$. This range encompasses the IPC flux 
(Feigelson et al. 1987) of
F$_{x}$(0.2-4 keV) = (3.9-4.6) $\times$ 10$^{-13}$ ergs cm$^{-2}$ s$^{-1}$.

To investigate in more detail the changes in X-ray emission that
occurred during ObsId 25693 we extracted time-partioned events
and spectra from the first $\approx$7 ks when the light curve
was nearly steady (pre-flare), the middle $\approx$8 ks when
the count rate rose and declined (flare), and the last $\approx$4 ks
(post-flare) when the count rate had declined to a value nearly the same 
as at the start of the observation. 
We use the term ``flare'' here in a general sense to denote the
rise and fall in count rate over a time interval of about 
two hours, even though the rise was not impulsive. Spectra
for each time segment were fitted and the results are summarized
in Table 2. 
Since the time-partitioned spectra contain fewer counts than
the total spectrum, uncertainties in best-fit parameters are larger.
Even so, the three segments show clear differences in 
temperature and flux.

At the start of ObsId 25693 the count rate, temperature,
and flux were already elevated above the low-state values
observed in the two subsequent observations.
A 2T $vapec$ fit of the pre-flare spectrum gives a hot 
component temperature kT$_{2}$ = 5.5 (3.4 - 12.7, 1$\sigma$ range) keV
with $\approx$47\% of the EM in the hot component.
The count rate and L$_{x}$ peaked during the flare and a 2T  $vapec$ 
fit yields kT$_{2}$ = 3.0 (2.4 - 4.3) keV with 
$\approx$61\% of the EM in the hot component and 
log L$_{x}$ = 30.17 ergs s$^{-1}$.  Even though the 
best-fit value of kT$_{2}$ during the flare segment was somewhat 
lower than during pre-flare, the uncertainty ranges are large enough
to allow overlap.
During the last $\approx$4 ks (post-flare) the emission had
fallen off to values only marginally higher than in low-state with
an isothermal 1T plasma model giving kT = 1.0 (0.9 - 1.1) keV.
The above results show that the event recorded in ObsId 25693 
was already in progress when the observation began but the 
emission had declined to near low-state levels by the end of the 
observation. The duration of the elevated emission (pre-flare$+$flare)
was at least $\approx$15 ks.

Since the elevated flux and temperature were not detected five days later at the 
same rotational phase in ObsId 25124, the high emission state is not
consistent with a long-lived active region on the star that might have 
been caught by chance rotating across the line-of-sight. But the 
observed X-ray variability in ObsId 25693 is consistent with a
modest X-ray flare.
The analysis of T Tauri star X-ray flares detected in a
XMM-Newton X-ray of the Taurus Molecular Cloud (TMC) by
Franciosini et al. (2007) revealed peak luminosities 
log L$_{x}$ $\geq$31 ergs s$^{-1}$, high plasma temperatures
$\geq$5 keV (60 MK), rise times of $\approx$1 - 35 ks, 
and decay timescales of $\approx$2 - 45 ks. The TAP 26 variability 
in Obs 25693 fits within these ranges. Even so, the additional
detection of slow low-amplitude variability in ObsId 25124 
when the emission was less luminous justifies further time monitoring 
to determine  whether the variability is modulated on longer 
timescales such as the planet's $\approx$10.8 day orbital period.

A previous XMM-Newton X-ray study of TTS in the TMC
by Telleschi et al. (2007) derived relations between L$_{x}$ and
stellar parameters. For weak-lined TTS they found a mean fractional
X-ray luminosity  log (L$_{x}$/L$_{*}$) = $-$3.36$\pm$0.07.
To compare this with TAP 26 we scale its value 
L$_{*}$/L$_{\odot}$ = 0.56 (0.45 - 0.71) as determined by Yu17 using d = 147 pc 
down to the {\em Gaia} DR3 distance of 121.6 pc which gives
L$_{*}$/L$_{\odot}$ = 0.38 (0.31 - 0.48).
The observed TAP 26 X-ray luminosites (Table 2) then give
log (L$_{x}$/L$_{*}$) = $-$3.36$\pm$0.10 (low-state) and
$-$2.96$\pm$0.10 (high-state). The low-state value thus
agrees well with the mean for weak-lined TTS in the TMC sample. 
But the high state value lies at the high end of the range
observed for weak-lined TTS with masses $>$0.7 M$_{\odot}$
and is atypical (Fig. 8-right of Telleschi et al. 2007).
Thus, the low-state value log L$_{x}$(0.3-8 keV) = 29.8 ergs s$^{-1}$ 
probably represents the typical X-ray luminosity of TAP 26.

\subsection{X-ray Ionization and Heating}

Using the L$_{x}$ values in Table 2 the unattenuated
X-ray flux at the planet's distance of 0.097 AU is
F$_{x,unatten}$ = (2.4 - 6.0) $\times$ 10$^{4}$ ergs cm$^{-2}$ s$^{-1}$.
This is $\sim$10$^{6}$ times greater than the Sun's flux at Jupiter
assuming a nominal solar X-ray luminosity log L$_{x,\odot}$ = 27.3 ergs s$^{-1}$. 

We estimate the X-ray ionization and heating rates at the 
TAP 26 planet's separation using the procedure given
for V830 Tau in Skinner \& G\"{u}del (2021) and
further details can be found therein.
The X-ray ionization rate at a distance $r$ from the star 
for a thermal plasma at  X-ray temperature kT$_{x}$ is (Shang et al. 2002)

\begin{equation}
\zeta(r) \approx \zeta_{\rm x} \left[{ \frac{r}{R_{\rm x}}} \right]^{-2} \left[{ \frac{kT_{x}}{\epsilon_{ion}}} \right] I(\tau_{\rm x})~~~\
({\rm s}^{-1}~{\rm per~ H~nucleus)}.
\end{equation}
where $\zeta_{\rm x}$ is the primary ionization rate (eq. [2]), $R_{x}$ is the distance of the 
X-ray source above the star, $\epsilon_{ion}$ $\approx$ 37 eV  is the energy to create an
ion pair, and $I(\tau_{\rm x}$) is an X-ray attenuation factor evaluated at optical
depth $\tau_{x}$. 

For coronal X-ray emission $R_{x}$ corresponds to at most a few stellar radii 
and we simply use $R_{x}$ $\approx$ R$_{*}$ = 1.17 R$_{\odot}$ (Yu17) for TAP 26,
so the computed value of $\zeta(r)$ will be a lower limit.
The results of Shang et al. (2002) include attenuation by a disk wind
in the expression for $I(\tau_{\rm x})$ but this effect is negligible for a 
weak-lined TTS like TAP 26 where the disk has dissipated so we adopt the approximate relation
$I(\tau_{\rm x}$) $\approx$ e$^{-\tau_{x}}$. The height $z$ in the planet's
atmosphere  corresponding to a given value $\tau_{x}$ depends on the atmospheric
model which explicitly specifies the number density of hydrogen nuclei versus height $n_{\rm H}$($z$).
Since the planet's properties are not yet characterized we do not adopt a specific 
planet atmospheric model. 
For a photon of energy $E$ the equivalent neutral hydrogen column density along 
the line-of-sight from the star to the point in the planet's atmosphere at 
which  $\tau_{x}$ = 1 is simply
N$_{\rm H,\tau_{x}=1}$ = 1/$\sigma(E)$. The X-ray photoelectric absorption cross-section declines
rapidly with photon energy as
$\sigma(E)$ = $\sigma_{0}$(E/1 keV)$^{-p}$ cm$^{-2}$
where $\sigma_{0}$ = 2.27 $\times$ 10$^{-22}$ cm$^{2}$ and
$p$ = 2.485 for solar abundances but the value of $p$ is abundance dependent
(Morrison \& McCammon 1983; Igea \& Glassgold 1999; Shang et al. 2002).
Thus lower energy X-rays are more heavily absorbed in the planet's outer atmosphere,
all other factors being equal. Scattering effects are negligible at energies below
a few keV of interest here (Bruderer et al. 2009).

The total X-ray ionization rate given by eq. (1) above accounts for multiple secondary 
ionizations resulting from the primary photoelectron ionization rate 

\begin{equation}
\zeta_{\rm x} =  1.13 \times 10^{-8}  \left[{ \frac{L_{x}}{10^{30}~ {\rm erg~ s}^{-1}}\
} \right] \left[{ \frac{kT_{\rm x}}{{\rm keV}}} \right]^{-(p+1)} \left[{ \frac{R_{\rm x}}{10^{12}~{\rm cm}}} \right]^{-2}~~~({\rm s}^{-1}).
\end{equation}

The X-ray heating rate per unit volume is

\begin{equation}
\Gamma_{\rm x} = \epsilon_{x}\zeta n_{\rm H} Q
\end{equation}
where 0 $<$ $\epsilon_{x}$ $<$ 1 is the fractional X-ray heating efficiency, 
$Q$ $\approx$ 20 eV is the heating rate per ionization and
$n_{\rm H}$ is the number density of hydrogen nuclei in the planet's
atmosphere at the height $z$ corresponding to the X-ray optical depth
at which $\zeta$ is computed (eq. 1). 
Rewriting the above equation in normalized form gives
\begin{equation}
\Gamma_{\rm x} = 3.2 \times 10^{-9} \epsilon_{x} \left[\frac{\zeta}{10^{-8}~{\rm s^{-1}~H^{-1}}}\right]\left[\frac{n_{\rm H}}{10^{10}~{\rm cm^{-3}}}\right]\left[\frac{Q}{20~\rm{eV}}\right]~~~{\rm (ergs~cm^{-3}~s^{-1}) } .
\end{equation}
Table 3 summarizes the ionization and heating rates for 
TAP 26 in low and high X-ray emission states evaluated at $\tau_{x}$ = 1.
Since we do not invoke a specific planet atmosphere model the
value of $n_{\rm H}$ used to compute $\Gamma_{\rm x}$ in Table 3 is left unspecified.
The high state heating rate in Table 3 is evaluated using the weighted plasma
temperature and the combined L$_{x}$ of the cool and hot plasma components,
as well as separately for the contributions of each temperature component. 
When the weighted temperature is used, the low and high state heating rates
are nearly the same. But when each temperature component is treated separately
the cool plasma dominates the heating. This clearly shows that the 
computed heating rate is sensitive to the assummed plasma temperature, a
consequence of the steep falloff in the primary photoionization rate 
$\zeta_{x}$ with increasing temperature (eq. [2]).

\begin{deluxetable}{lccccccc}
\tabletypesize{\footnotesize}
\tablewidth{0pt}
\tablecaption{X-ray Ionization and Heating Rates (TAP 26)}
\tablehead{
           \colhead{State}           &
           \colhead{r}               &
           \colhead{kT$_{x}$}        &
           \colhead{log L$_{x}$}     &
           \colhead{$\zeta_{x}$}     &
           \colhead{$\zeta(r)$}      &
           \colhead{N$_{\rm H}$($\tau_{x}$=1)}      &
           \colhead{$\Gamma_{x}$}      \\
           \colhead{}                &
           \colhead{(au)}            &
           \colhead{(keV)}           &
           \colhead{(ergs s$^{-1}$)}           &
           \colhead{(s$^{-1}$)}      &
           \colhead{(s$^{-1}$ H$^{-1}$)}   &
           \colhead{(cm$^{-2}$)}                &
           \colhead{(ergs s$^{-1}$ cm$^{-3}$ $n_{\rm H}^{-1}$)} 
 }
\startdata
\vspace{0.05in}
 Low\tablenotemark{a}   & 0.0968 & 1.02            & 29.8  & 1.00e-06   & 3.2e-08  & 8.45e21   & 1.0e-18  \\ 
 High\tablenotemark{b}  & 0.0968 & 1.62            & 30.2  & 0.50e-06   & 2.6e-08  & 1.21e22   & 8.3e-19  \\
 High\tablenotemark{c}  & 0.0968 & 0.38 (kT$_{1}$) & 29.94 & 4.32e-05   & 51.4e-08 & 4.52e20   & 1.6e-17 \\
 High\tablenotemark{c}  & 0.0968 & 2.9 (kT$_{2}$) & 29.85 & 2.96e-08   & 0.3e-08  & 6.21e22   & 8.6e-20 \\
\enddata
\tablecomments{
The  secondary ionization rate $\zeta$ (eq. [1]), 
primary ionization rate $\zeta_{x}$ (eq. [2]), and
heating rate $\Gamma_{x}$ (eq. [3]) are computed 
at $\tau_{x}$ = 1 using $\epsilon_{x}$ = 1,
R$_{x}$ = R$_{*}$ = 1.17 R$_{\odot}$, Q = 20 eV, and the assumed planet separation $r$ = $a$ = 0.0968 au.
N$_{\rm H}$($\tau_{x}$=1) = 1/$\sigma$(E$_{50}$) where E$_{50}$ is the 
median photon energy (Table 2).
The number densiity  $n_{\rm H}$ at height $z$ in the planet's atmosphere corresponding
to $\tau_{x}$=1 required to evaluate $\Gamma_{x}$ depends on the
adopted planet atmosphere model and has been left as a free parameter.}
\tablenotetext{a}{Low state values of kT$_{x}$ and L$_{x}$ are from the 1T $apec$
                  fit of ObsIds 25124 and 26225 (Table 2).}
\tablenotetext{b}{High state values of kT$_{x}$ (weighted) and L$_{x}$ are 
                  from the 2T $vapec$ fit of ObsId 25693 (Table 2),}
\tablenotetext{c}{High state values are evaluated separately for the cool (kT$_{1}$) and 
                  hot (kT$_{2}$) plasma components from the 2T $vapec$ fit of ObsId 25693 (Table 2),}
\end{deluxetable}
\clearpage

\clearpage

\subsection{EUV Heating}

The planet's outer atmosphere will also be ionized and heated by
stellar EUV photons ($\lambda$ = 124 - 920 \AA, E =  0.013 - 0.1 keV).
The heating is mitigated by competing processes such as 
radiative cooling by the far-UV Ly$\alpha$ line at 
1215.7 \AA~ (Salz et al. 2016).
We apply to TAP 26 the same procedure as was used for V830 Tau
(Skinner \& G\"{u}del 2021) to estimate the EUV heating rate.
Taking L$_{x}$ $\approx$ 10$^{30}$ ergs s$^{-1}$ as a representative
value for TAP 26 the heating rate for a monoenergetic EUV photon flux at energy 
E at the planet's separation 0.097 au is, by analogy
with eq. [6] of Skinner \& G\"{u}del (2021) 
\begin{equation}
\Gamma_{\rm EUV} = 3.8 \times 10^{4} \eta \left[\frac{L_{EUV}}{L_{x}}\right] \sigma(E) {\rm e}^{-\tau_{\rm euv}}  n_{\rm{H}}~~~{\rm (ergs~cm^{-3}~s^{-1}) } .
\end{equation}
In the above, 0 $<$ $\eta$ $<$ 1 is the fractional EUV heating efficiency and the EUV cross-section is 
approximated by
$\sigma$(E) = 6 $\times$ 10$^{-20}$(E/100 eV)$^{-p}$ (cm$^{2}$) with $p$ = 2.485 (Bruderer et al. 2009).
As for X-rays $n_{\rm H}$ is the H number density at height $z$ corresponding to $\tau_{euv}$.
At a characteristic EUV energy E$_{0}$ = 60 eV  unit optical depth 
$\tau_{euv}$ = 1 occurs at N$_{\rm H}$ = 5 $\times$ 10$^{18}$ cm$^{-2}$,
about 3 orders of magnitude less than $\tau_{x}$ = 1 at 1 keV. Thus,
X-ray photons will penetrate to deeper layers of the planet's atmosphere.
Inserting values for $\sigma(E)$ at
E = 60 eV and $\tau_{euv}$ = 1 in the above expression and 
normalizing the H number density to a representative hot Jupiter value 
(Penz et al. 2008; Murray-Clay et al. 2009, hereafter MC09) gives at $a$ = 0.097 AU
\begin{equation}
\Gamma_{\rm EUV} = 3 \times 10^{-5} \eta \left[\frac{L_{EUV}}{L_{x}}\right] \left[\frac{n_{\rm H}}{10^{10}~{\rm cm^{-3}}}\right] ~~~{\rm (ergs~cm^{-3}~s^{-1}) } .
\end{equation}
Heating efficiencies $\eta$ $\approx$ 0.1 - 0.6 have been used in theoretical
studies (e.g. Penz et al. 2008; Shaikhislamov et al. 2014).
L$_{EUV}$ is generally not measurable from direct observations
due to interstellar absorption but the approximation
L$_{\rm EUV}$ $\approx$ L$_{x}$ is sometimes adopted (e.g. Ribas et al. 2005; Owen \& Jackson 2012).
The evolutionary tracks derived by Tu et al. (2015) predict L$_{\rm EUV}$ and L$_{x}$ as a
function of age and an assumed stellar rotation rate at an age of 1 Myr.
At an age of $\sim$17 Myr the predicted values of L$_{\rm EUV}$ and L$_{x}$
are similar to within a factor of a few.

\subsection{Implications of XUV Irradiation of the Planet's Atmosphere}

Although the planet's orbital separation and mass have been estimated its 
radius and atmospheric properties are unknown. In general, the atmospheres of
hot Jupiters are not yet well-characterized so any discussion
of XUV effects is subject to uncertainties in the adopted atmospheric model. 
Given our lack of knowledge of the atmospheric properties of TAP 26 b,
detailed modeling is not yet warranted but some insight can be obtained 
based on previous studies of other hot Jupiters.
 
Hot Jupiters at close separation from TTS will experience 
photoevaporative mass loss. Since TTS are generally strong
XUV sources, several studies have concluded that the planet's
mass-loss rate should be determined using numerical 
hydrodynamic models (e.g. Yelle 2004; Penz et al. 2008; MC09;
Salz et al. 2016). But other studies using analytic 
energy-limited approach predict mass loss rates
that are not much different than obtained using hydrodynamic
models (e.g. Shaikhislamov et al. 2014).

Hydrodynamic models assume a steady spherical wind similar to that of 
the solar wind developed by Parker (1958; 1965), except externally heated
(ignoring any internal heating such as may arise from tidal dissipation).
The mass loss rate as a function of radial distance $r$ from the planet
is $\dot{M}_{p} = 4\pi r^{2} \rho(r) v(r)$ where
$\rho(r)$ and $v(r)$ are the the wind's mass density and speed.
Under the simplifying assumption that the wind is {\em isothermal},
the wind speed equals the sound speed $c_{s}$ at the critical point 
distance $r_{s}$ = GM$_{p}$/(2$c_{s}^2$); in this case the crtical
point occurs at the sonic point.  The isothermal sound speed is
$c_{s}$ = ($\mathcal{R}$T/$\mu$)$^{1/2}$ where $\mathcal{R}$
is the gas constant and $\mu$ is the mean weight per particle (amu)
in the wind. For a pure neutral hydrogen atmosphere $\mu$ = 1 and
if fully ionized $\mu$ = 0.5. Even though simulations of hot
Jupiter atmospheres indicate the wind is not strictly isothermal
(e.g Salz et al. 2016), the isothermal approximation is still
useful as a gauge of the mass loss rate, especially in view of
all the other uncertainties involved such as atmospheric abundances.

Denoting the mass density at the sonic point 
as $\rho_{s}$ gives the normalized mass loss rate in the isothermal approximation
\begin{equation}
\dot{M}_{p,s} \sim 10^{-9}\left[\frac{r_{s}}{10~{\rm R_{J}}}\right]^2\left[\frac{\rho_{s}}{10^{-15}~ {\rm g~cm^{-3}}}\right]\left[\frac{c_{s}}{10~ {\rm km~s^{-1}}}\right]~~~{\rm (M_{J}~ yr^{-1}) }.
\end{equation}
 
Although the flux F$_{XUV}$ incident on the planet does not appear explicitly 
in the above, it enters implicitly since it affects the temperature profile T($r$) 
via heat deposition and the quantities which determine the mass loss rate 
are temperature dependent.
Detailed atmospheric radiative transfer modeling taking both heating and cooling into account
are needed to reconstruct T($r$). Hydrodynamic models for hot Jupiters 
orbiting TTS with strong XUV emission predict that the atmospheric
temperature will increase rapidly from a base level T$_{0}$ to a maximum
value T$_{max}$ and then slowly decline outward
(Penz et al. 2008; MC09; Salz et al. 2016; Shaikhislamov et al. 2014).
The maximum temperature is expected to self-regulate or ``thermostat'' 
at T$_{max}$ $\sim$ 10,000 K as a result of XUV heating being offset by
radiative and adiabatic cooling.
However, recent observations of the hot Jupiter WASP-189b suggest somewhat
higher temperatures are possible (Sreejith et al. 2023). 
The base temperature T$_{0}$ can be approximated by the planet's 
equilibrium temperature T$_{eq}$. Adopting an effective temperature
T$_{eff,*}$ = 4620 K for TAP 26 (Yu17) gives T$_{eq}$ = 760 K for TAP 26 b,
assuming a Bond albedo A$_{B}$ $\approx$ 0.05 - 0.1 (Sudarsky et al. 2000)
with heat evenly distributed over the planet's surface.
Simulations of hot Jupiter atmospheres typically predict the flow to 
become supersonic at $r_{s}$ $\approx$ 3 - 5 R$_{p}$ 
(MC09; Owen \& Adams 2014; Salz et al. 2016). 
At the sonic point the temperature is predicted to be a few thousand K.
At T $\sim$ 2,000 - 4,000 K the sound speed in neutral H
gas is $c_{s}$ = 4 - 6 km s$^{-1}$.

In order to calculate the XUV heating rate  (eqs. [4], [6]) 
the number density $n_{\rm H}$ is needed, and it varies with 
height in the planet's atmosphere.
Hot Jupiter models usually assume a base level density
(or pressure) and extrapolate it outward.
The base level density (or pressure) is quite uncertain and its 
uncertainty propagates into the derived mass 
loss rate (Shaikhislamov et al. 2014; Salz et al. 2016). 
Hot Jupiter simulations
predict a typical number density at the sonic point
n$_{\rm H.s}$ $\sim$ 10$^{6}$ - 10$^{7}$ cm$^{-3}$,
or an equivalent neutral hydrogen mass density 
$\rho_{s}$ = $\mu$m$_{\rm H}$n$_{\rm H,s}$ $\sim$ 10$^{-18}$ - 10$^{-17}$ g cm$^{-3}$.
Conversion of number to mass density requires knowledge of
the atmsophere's chemical composition which is generally
not well-known and can vary with height, but  
observational constraints have been obtained for a few objects
(e.g. Charbonneau et al. 2002; Cody \& Sasselov 2002; see also
the review of Fortney et al. 2021). 

Using the above density estimates as a guide in eq. (7)
and assuming M$_{p}$ $\approx$ 2 M$_{J}$ (Yu17) 
and  radius R$_{p}$ $\approx$ 1.3 R$_{J}$ (Sarkis et al. 2021)
for TAP 26 b, the inferred mass loss rate is   
$\dot{M}_{p}$ $\ltsimeq$ 10$^{-11}$ M$_{\rm J}$ yr$^{-1}$.
Previous studies of late-type stars predict that most of the
exoplanet's photoevaporative mass-loss will occur during
the first Gyr (Jackson et al. 2012). If the mas-loss rate
of TAP 26 b is set equal to the above upper limit and assumed
to be constant over time, the hot Jupiter would shed less than 
1\% of its mass during the first Gyr of its life.
This conclusion is strengthened when taking into account that
the star's  XUV flux will decline with age (G\"{u}del et al. 1997)
along with the  XUV-driven mass loss rate.

For comparison with the above, the energy-limited mass-loss rate is
(Sanz-Forcada et al. 2011)
$\dot{M}_{p,el}$ = 3F$_{\rm XUV}$/(4G$\rho_{p}\mathcal{K})$ where
$\rho_{p}$ is the planet's mean mass density, 0 $<$ $\mathcal{K}$ $\leq$ 1
accounts for Roche lobe effects, G is the gravitational constant and
F$_{\rm XUV}$ is the XUV flux at the planet's surface.  
Assuming Roche lobe effects
are negligible ($\mathcal{K}$ $\approx$ 1; Erkaev et al. 2007; Lammer et al. 2009) 
and normalizing to Jupiter's mass density $\rho_{J}$ = 1.3 g cm$^{-3}$ gives
$\dot{M}_{p,el}$ $\sim$ 10$^{-16}$($\rho_{p}$/$\rho_{J}$)$^{-1}$F$_{\rm XUV}$ (M$_{J}$ yr$^{-1}$).
Expressing unattenuated high-energy flux F$_{\rm XUV}$ in terms
of the star's XUV luminosity L$_{\rm XUV}$ = L$_{x}$ $+$ L$_{\rm EUV}$ and normalizing gives 
\begin{equation}
\dot{M}_{p,el} \approx 5\times10^{-12}\left[\frac{\rho_{p}}{\rho_{\rm J}}\right]^{-1}\left[\frac{a}{0.1~{\rm au}}\right]^{-2}\left[\frac{L_{\rm XUV}}{10^{30}~ {\rm erg~ s^{-1}}}\right]~~(M_{\rm Jup}~{\rm yr}^{-1}).
\end{equation}
Adopting L$_{x}$ = 10$^{30}$ erg s$^{-1}$ for TAP 26 and assuming L$_{\rm EUV}$ $\sim$ L$_{x}$ yields
$\dot{M}_{p,el}$ $\sim$ 10$^{-11}$($\rho_{p}$/$\rho_{J}$)$^{-1}$~ M$_{J}$ yr$^{-1}$.
This energy-limited mass-loss rate is the same order of magnitude as the upper limit estimated 
above using the hydrodynamic approach but is subject to the aforementioned uncertainty in L$_{\rm EUV}$.

The above estimates do not include the effects on mass loss due to any planetary
magnetic field. The degree to which the planet's wind is controlled by the 
B-field depends on the ratio of the wind's ram pressure to the magnetic pressure
and a modest field strength B$_{p}$ $\sim$ 0.3 G can lead to a significant 
reduction in $\dot{M}_{p}$ (Owen \& Adams 2014).

\subsection{Comparison with V830 Tau}

The solar-mass weak-lined TTS V830 Tau is remarkably similar to TAP 26
and may also host a close-in hot Jupiter V830 Tau b (Donati et al. 2015, 2016, 2017).
But an attempt to confirm the radial velocity (RV) detection of the planet
by Damasso et al (2020) yielded negative results so the planet's
existence needs confirmation. The RV data yielded an orbital period
P$_{orb}$ = 4.927 d and separation $a$ = 0.057 au, placing V830 Tau b
in a tighter orbit than TAP 26 b.
The K7-type host star V830 Tau has an estimated age of 2.2 Myr (Donati et al. 2015)
and a 2.74 d rotation period (Grankin et al. 2008). It may thus be a
younger but more slowly rotating analog of TAP 26, as discussed by Yu17.

Chandra observations of V830 Tau revealed variable X-ray emission
with log L$_{x}$ = 30.1 - 30.4 erg s$^{-1}$ (Skinner \& G\"{u}del 2021).
But XMM-Newton observations recorded brighter flare emission up to
log L$_{x}$ = 30.87 erg s$^{-1}$ (Franciosini et al. 2007,
adjusted to d = 130.4 pc). The lower end of the V830 Tau
range overlaps that of TAP 26 but during flares V830 Tau
is much more luminous than observed so far for TAP 26.
Since X-ray time monitoring is still quite limited for both
stars the full range of X-ray variability may not yet be sampled.
At their young ages the Tu et al. (2015) evolutionary
tracks place V830 Tau in the X-ray saturation regime and TAP 26
also in or near saturation. The mean hardness ratio
H.R = 0.36 and median photon energy E$_{50}$ = 1.60 keV of V830 Tau are
greater than TAP 26 (Table 2) even though both stars  have similar A$_{\rm V}$.
Thus, V830 Tau is characterized by hotter X-ray plasma.
Also, the putative hot Jupiter orbiting V830 Tau lies at a
separation $a$ = 0.057 au, half that of TAP 26 b. Thus,
the unattenuated X-ray flux at V830 Tau b is at least
four times greater than that at TAP 26 b.

\clearpage

\section{Summary}

\begin{enumerate}
      
\item New observations of TAP 26 reveal a variable
      X-ray source whose X-ray flux and temperature changed
      over a few hours with temperature of the hotter plasma 
      increasing to kT$_{x}$ $\approx$ 4 keV.  Slower variability 
      is also present.
      The X-ray flux range measured by Chandra
      brackets the flux measured previously by the Einstein 
      Observatory. The Chandra X-ray luminosity is in the 
      range log L$_{x}$(0.3-8 keV) =  29.80 - 30.20 ergs s$^{-1}$.
      
\item Since the elevated X-ray emission detected on 1 Dec. 2021
      was not present five days later at about the same rotation
      phase, the high X-ray state is evidently not tied to stellar
      rotation. Time coverage is not yet sufficient to determine
      whether the slower X-ray variability is modulated at the planet's orbital period.

\item The unattenuated X-ray flux impinging on the planet is
      F$_{x,unatten}$(0.3-8 keV) = (2.4 - 6.0)$\times$10$^{4}$ erg cm$^{-2}$ s$^{-1}$,
      or $\sim$10$^{6}$ times greater than the Sun's X-ray flux at Jupiter.
      The X-ray heating rate of the planet's atmosphere 
      is $\Gamma_{x}$ $\sim$ 10$^{-18}$ 
      ergs s$^{-1}$ cm$^{-3}$ $n_{\rm H}^{-1}$ where 
      the H-nucleus number density $n_{\rm H}$ at
      the atmospheric height corresponding to $\tau_{x}$=1
      is model-dependent and unspecified.

\item A comparison of TAP 26 b with previous studies of other hot
      Jupiters indicates that photoevaporative mass loss will
      be negligible over the first Gyr of the planet's life.

\end{enumerate}

\clearpage

\begin{acknowledgments}
Support for this work was provided by  Chandra award 
number GO2-23009X issued by the Chandra X-ray Center, which is operated by
the Smithsonian Astrophysical Observatory (SAO) for and on behalf of NASA.
This work has utilized HEASOFT developed and maintained by HEASARC at NASA GSFC.

\noindent This paper employs a list of {\em Chandra} datasets obtained by the
Chandra X-ray Observatory contained in
~\dataset[DOI: 10.25574/cdc.218]{https://doi.org/10.25574/cdc.218}.
\end{acknowledgments}

\vspace{5mm}
\facilities{CXO}

\vspace{5mm}
\software{CIAO (Fruscione et al. 2006),
          XSPEC (Arnaud 1996)}

\clearpage

\end{document}